\documentclass[fleqn,usenatbib]{mnras}

\usepackage{newtxtext,newtxmath}

\usepackage[T1]{fontenc}
\usepackage{ae,aecompl}
\usepackage{listings}

\usepackage{graphicx}	
\usepackage{amsmath}	
\usepackage{amssymb}	
\usepackage{soul}

\newcommand{\vphi}{$v_{\phi}$}
\newcommand{\alphaH}{$[\alpha/\mathrm{Fe}]$}

\title[The tale of the tail]{The tale of the tail - disentangling the high transverse velocity stars in Gaia DR2}

\author[Amarante et al.]{Jo\~ao A. S. Amarante,$^{1}$\thanks{E-mail: joaoant@gmail.com}
Martin C. Smith,$^{1}$\thanks{E-mail: dr.mcsmith@me.com}
Corrado Boeche$^{2}$
\\
$^{1}$Key Laboratory for Research in Galaxies and Cosmology\\ Shanghai  Astronomical Observatory, Chinese Academy of Sciences\\ 80 Nandan Road, Shanghai 200030, China\\
$^{2}$ INAF-Osservatorio Astronomico di Padova, vicolo dell'Osservatorio 5, 35122 Padova, Italy
\\
}

\date{Accepted XXX. Received YYY; in original form ZZZ}

\pubyear{2019}
\begin{document}
\label{firstpage}
\pagerange{\pageref{firstpage}--\pageref{lastpage}}
\maketitle

\begin{abstract}
Although the stellar halo accounts for just $\sim$~1\% of the total stellar mass of the Milky Way, the kinematics of halo stars encode valuable information about the origins and evolution of our Galaxy. It has been shown that the high transverse velocity stars in Gaia DR2 reveal a double sequence in the Hertzsprung-Russell (HR) diagram, indicating a bifurcation in the local stellar halo within 1 kpc. We fit these stars by updating the popular Besan\c con/Galaxia model, incorporating the latest observational results for the stellar halo and an improved kinematic description for the thick-disk from \citet{schon2012}. We are able to obtain a good match to the Gaia data and provide new constraints on the properties of the Galactic disc and stellar halo. In particular, we show that the kinematically defined thick disc contribution to this high velocity tail is $\approx 13\%$. We look in greater detail using chemistry from LAMOST DR5, identifying a population of retrograde stars with thick-disc chemistry. Our thick disc kinematic model cannot account for this population and so we conclude there is likely to be a contribution from heated or accreted stars in the Solar Neighbourhood.
We also investigate proposed dynamical substructures in this sample, concluding that they are probably due to resonant orbits rather than accreted populations. Finally we provide new insights on the nature of the two sequences and their relation with past accretion events and the primordial Galactic disc.\end{abstract}




\section{Introduction}\label{sec:intro}

The study of the stellar halo provides valuable information about our Galaxy's formation. It is especially important for learning about the early stages, since the halo contains most of the metal-poor (and hence oldest) stars in the Galaxy.
However the difficulty in studying the stellar halo is the relative paucity of stars compared with the dominant population of disc stars. Its contribution to the total stellar mass of the Galaxy is roughly 1\%, while the disc contributes approximately $\sim$ 90\% \citep{binney}. Closer to solar neighbourhood, e.g. within 1 kpc from the Sun, the stellar halo fraction is even lower, with values ranging from 0.1 to 1\%. \par 
In the last decade, several studies have provided clues to the local stellar halo fraction, $f_h$, using different methods and volumes\footnote{Throughout this paper, we take $f_h$ and $f_{TD}$ to refer to the fraction of halo and thick disc stars, respectively, within 1 kpc of the Sun, e.g. $f_h$ is the number of halo stars divided by the total number of halo, thick and thin disc stars. In the rare instances that we refer to the fraction at the Solar position (i.e. the in-plane normalisation at the Solar radius), we denote this as $f_{h,\odot}$ or $f_{TD,\odot}$.}.
For example, \citet{gould} used deep HST star counts to constrain $f_{h,\odot}$, finding it to be in the range 0.07 to 0.2\%.
Later, \citet{juric}, using the large photometric Sloan Digital Sky Survey (SDSS), modelled the number density of stars in the Milky Way and found that, at the Solar position, $f_{h,\odot} = 0.4$\%. Large spectroscopic surveys, such as the Radial Velocity Experiment (RAVE), also reinvigorated studies of the Milky Way's structure. \citet{kordopatis}, using RAVE DR4 data, found that the local $f_h$ varies from 0.5\% to 0.6\% depending on whether they use a chemical or kinematic selection, respectively. More recently, \citet{posti} used the cross-match between Gaia DR1 and RAVE to select, through Action Angle distributions \citep{binney}, the local stellar halo. From their study they conclude that $f_h \approx 0.45$\%. 
Even though these studies all use different selection criteria and could be subject to various selection effects, it seems that the value of $f_h$ within 1 kpc seems to be converging on $\approx 0.5$\%.

The value of $f_h$ is an important ingredient for models of the distribution of stars in our Galaxy. The most influential model in the past two decades has been the Besan\c con model. In their original model, which was the standard model for most of this time, \citet{robin} adopted a value for the halo normalisation based on star counts at high and medium latitudes. They estimated $f_{h,\odot} = 0.06$\% for $M_V<8$, which corresponds to $f_h \approx 0.06$\%. This was subsequently updated in \citet{robin2014} to incorporate a new multi-component thick disc, but still a low value for the halo normalisation was adopted ($f_h = 0.04$). Therefore it is worth emphasising that the halo normalisation in the Besan\c con model is significantly less than the above consensus. This means that models and data products which are based on the Besan\c con model, such as the Galaxia model or Rybizski mock Gaia catalogue, will significantly underestimate the halo contribution.

Since the thick disc was first proposed as a distinct component of the Milky Way, its local fraction, $f_{TD}$, is still an open question: due to the degeneracy with its scale length, $f_{TD}$ ranges from 2\% \citep{gilmore} to 12\% \citep{juric}. We refer to \citet{bland} review where they found a mean value of $f_{TD}=4\%\pm2\%$, after compiling 25 results from photometric surveys without any detailed stellar abundance information. Nonetheless, it's worth mentioning a few specific studies, e.g., \citet{kordopatis} found that the local $f_{TD}$ was either 15.2\% or 18.3\%, depending on whether a kinematic or chemical selection was used, respectively. The caveat with these values is the assumption of a Gaussian distribution for both azimuthal velocity, $v_{\phi}$, and metallicity distributions. As will be discussed later, the asymmetric drift in the disc makes the \vphi\, distribution skew towards lower velocities, thus a Gaussian may not be physically meaningful to describe it. Moreover, it may underestimate (\textit{overestimate}) the contribution of the thin (\textit{thick}) disc stars. 
Finally, if high resolution spectroscopic data are included and the thin and thick discs are defined as [$\alpha$/Fe] poor and rich, respectively, the scale length and scale heights are clearly different for both populations \citep{bensby}. For example, \citet{bovy}, find that the [$\alpha$/Fe] rich stars have a larger scale height and smaller scale length than the [$\alpha$/Fe] poor stars. This helps to break the aforementioned degeneracy with $f_{TD}$ and favour lower values for the thick disc normalisation. \par
Another challenge is to correctly distinguish stellar halo and thick disc stars when dealing with a local sample. There is a non-negligible overlap between the former's metal rich end and the latter's metal poor end, with both co-existing at around [Fe/H]$\approx -0.8$, which means that some cuts in [Fe/H] could lead to contaminated samples. Nevertheless, $\alpha$ peak elements, especially magnesium, can help distinguishing both populations. For example, using data from APOGEE \citet{hawkins} and \citet{hayes} have shown that for [Fe/H] $\lesssim-0.7$ there are two distinct populations: an \alphaH\, poor sequence with zero net rotation, which is consistent with stars being mainly accreted from small dwarf satellites whose stellar populations were formed in the first few Gyr of the universe; and an \alphaH\, rich sequence, which has some net rotation and is kinematically colder than the \alphaH\, poor sequence.
\citet{hawkins} point out that these \alphaH\, rich stars could come from either the thick disc or from some form of in-situ (in their words "canonical") halo population. They conclude that these two populations are not chemically distinct and that this may suggest a common origin.

\citet{gaiahr} have demonstrated the amount of detail it's possible to obtain from the analysis of the HR diagrams drawn from Gaia data \citep{gaia1, gaia2}. For instance, for those interested in the stellar halo and thick disc, their figure 21 is especially relevant (reproduced here in Section \ref{sec:chem}). By selecting stars with transverse velocity greater than 200 km/s, two distinct sequences are clearly distinguishable in the colour-magnitude diagram. This kinematic selection cut is likely to produce a sample dominated by halo stars and so the fact that two sequences are observed could be an indication that the local halo is composed of two different stellar populations. However, one might wonder whether this cut produces a pure halo sample. Could there be contamination from the thick disc and, if so, how does this affect our interpretation of the data?
For that matter, the relative fractions of the stellar halo and the thick disc at the Solar Neighbourhood, turn out to be a very important constrain, nonetheless not trivial to obtain.
\par
In this paper, we present a different approach to estimate $f_h$ and $f_{TD}$ in order to quantify the amount of thick disc contamination in the Gaia high transverse velocity sample. As discussed above, this could potentially change the interpretation of a dual stellar halo population in the solar neighbourhood, as suggested by \citet{haywood} and \citet{gallart}. The paper is organised as follows: Section \ref{sec:data} describes the data we will use in our analysis; Section \ref{sec:model} models the tail of the high transverse velocity distribution, in order to obtain $f_h$ and $f_{TD}$; Section \ref{sec:chem} dissects the chemistry of these stars, using spectroscopic data from LAMOST; Section \ref{sec:dyn} explores the dynamical properties of the sample; Section \ref{sec:summary} brings together our findings and presents our take on the nature of the stars in this sample; finally in Section \ref{sec:conclusion} we provide some brief conclusions. We also include two appendices: firstly presenting the ADQL query we used to obtain our Gaia sample (Appendix \ref{sec:adql}); and secondly investigating proposed substructures in the space of orbital apocentre and maximum height from the plane (Appendix \ref{sec:apo-zmax}).

\section{Data selection}\label{sec:data}
The sample was extracted from the GDR2 catalogue using the same ADQL query as described in \citet{gaiahr}, which we reproduce in Appendix \ref{sec:adql}. The query selects all stars 
within 1 kpc that have G $<17$ and with transverse velocity $v_t = 4.74/\varpi\sqrt{\mu^2_{\alpha^*} + \mu^2_{\delta}}> 200$ kms/s.
After some additional quality cuts (again described in Appendix \ref{sec:adql}), the total sample size is 77,107. In addition to this sample
Since in Section \ref{sec:model} we also wish to compare to the entire $v_t$ distribution, we randomly selected a sample of 672,138 stars that satisfy the same constraints as the high $v_t$ main sample except without the $v_t$ cut. \par
Throughout the paper, we use the Galactocentric cylindrical velocities $v_R$, $v_{\phi}$ and $v_z$, where they are positive in the direction to the Galactic centre, Galactic rotation and North Galactic pole, respectively. We adopt from \citet{schon2012akin} the values for the solar Galactocentric distance $R_{\odot} = 8.27$ kpc and local circular velocity $v_c = 238$ km/s. We assume the Solar motion with respect to the Local Standard of Rest to be $(U_{\odot}, V_{\odot}, W_{\odot})=(11.1, 12.24, 7.25)$ km/s \citep{schon2010}. %
\section{Transverse velocity modelling}\label{sec:model}
\subsection{Galaxia: a synthetic population model}
The Galaxia code\footnote{\url{http://galaxia.sourceforge.net/}} \citep{sharma} is a population synthesis model based on the Besan\c con Galaxy model \citep{robin}, which is itself constrained using a variety of observational results.
Galaxia is a convenient tool for understanding large observational surveys, such as Gaia, and is especially useful for modelling selection effects in all-sky surveys. In Galaxia, both the stellar halo and thick disc have their 3D velocities described by a Gaussian distribution. We summarise the parameters of each population in Table \ref{tab_param}, where these parameters are the same as \citet{robin}.
It is important to note that the Besan\c con model used in Galaxia is the original model, as described in \citet{robin}, and not the updated model presented in \citet{robin2014} and available online\footnote{\url{http://model2016.obs-besancon.fr/}}. This newer model, which incorporates a  multi-component thick disc, does not significantly improve the match to the $v_t$ distribution, as we will discuss later in this section.

\begin{table}
 \label{tab_param}
 \begin{center}
 \begin{tabular}{ccc|cc}
 \hline
  & \multicolumn{2}{c|}{\textbf{Galaxia}}&\multicolumn{2}{c}{\textbf{Model}}\\
  \cline{2-3} \cline{4-5}\\

  & TD & Halo & TD & Halo \\
  \hline
  $\langle v_R \rangle \,; \, \sigma_{v_R}$ [km/s] & $[0;66]$ & $[0,135]$ &[$0;58]$ & $[0,140]$ \\
$\langle v_{\phi} \rangle \,; \, \sigma_{v_{\phi}}$ [km/s] & $[195;51]$ & $[11;85]$ &[$196,48]^*$ & $[30,78]$ \\
  $\langle v_z \rangle \,; \, \sigma_{v_z}$ [km/s]  & $[0;42]$ & $[0,84]$ &[$0;42]$ & $[0,80]$ \\
  $f_{pop}(\varpi > 1$) [\%] & 10.95 & 0.11 & 6.52 & 0.47\\ 
 \hline
 \end{tabular}
  \caption{Parameters for the canonical Galaxia model and our updated model. \newline * We do not use a Gaussian distribution to model $v_\phi$ for the thick disc and so these numbers are just included to give an indication of the mean and spread of our distribution. See text for details.}
  \end{center}
\end{table}
In order to understand the properties of our observed Gaia sample, a good starting point is to compare what would be the expected transverse velocity distribution from a canonical model of the Milky Way. For this purpose, we use the mock catalogue created by \citet{mock}, available at GAVO\footnote{\url{http://dc.g-vo.org/tap}} as \textit{gdr2mock.main}, which is based on the Galaxia canonical model. Using the same ADQL query as for the observational sample, a total of 42,128 mock stars with $v_t > 200$ km/s were obtained.\par
\begin{figure}
    \includegraphics[width=\columnwidth]{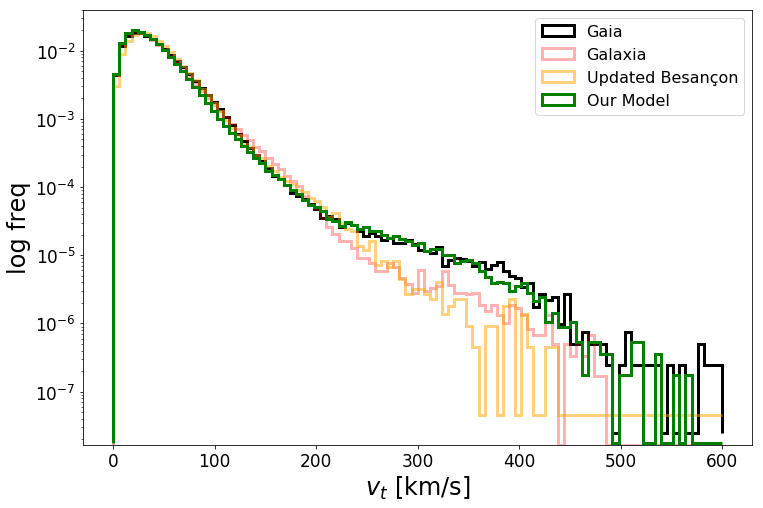}
    \caption{The normalised histogram of $v_t$ for Gaia data and our model, in black and green respectively. The canonical Galaxia model, in red, is unable to describe the observed data for $v_t>200$ km/s mainly due to the adopted normalisation factors for the thick disc and stellar halo populations. The updated Besan\c con model \citep{robin2014}, in yellow, is also unable to fit the tail of the $v_t$ distribution.} 
    \label{vts}
\end{figure}

Although the full mock catalogue has the same number of stars as the Gaia DR2 catalogue, it is not expected to precisely match the number of stars in either a specific stellar population (e.g. thick disc or halo) nor a selected sample (e.g. $v_t$ > 200 km/s). For example, even though the total number of stars in the mock and observed samples are similar for the same magnitude and spatial cuts, namely 25 and 29 million stars, respectively, if we look at the stars with $v_t>200$ km/s we find that the numbers are 42 and 80 thousand stars, respectively. The factor of two difference already indicates that there may be a deficiency in the underlying model from which the mock sample is drawn. This is shown more clearly in Fig \ref{vts}, where we have normalised both distributions by the total number of stars that satisfy the original ADQL query without the $v_t$ cut, i.e. over the whole range of transverse velocities. Clearly, the canonical model of the Galaxy is unable to reproduce the observed data as there is a significant lack of stars for $vt\gtrsim200$ km/s. For completeness we also include the distribution from the updated Besan\c con model \citet{robin2014}. Although this marginally improves the fit for moderate values of $v_t$, it still significantly underestimates the number of stars in the tail beyond $v_t \approx 250$ km/s.

The fact that these distributions don't agree could be due to either (a) how the velocities are assigned to each population within a given model, or (b) the assumed normalisation fraction of each population. In the high $v_t$ tail the two dominant populations are the stellar halo and thick disc. For the rest of this section we will focus on these two populations, describing our prescription for obtaining an improved parameterisation of the thick disc and halo.

\subsection{Thick disc}\label{sec:modelTD}
It is well known that the asymmetric drift affects the $v_{\phi}$ distribution of disc stars, meaning that a single Gaussian cannot provide a good match to the skewed observed distribution (e.g. \citealt{binney}). One common approach is to model this with two Gaussians, each with different mean ($\mu$), standard deviation ($\sigma$) and weight ($w$). For example, 
in their study of stars between 1 and 2 kpc above the Galactic plane, \citet{kordopatis} decomposed the observed $v_{\phi}$ distribution using three Gaussians, each one representing the thin disc, the thick disc and the halo. Although the results are consistent with the expected fractions for thin and thick disc, the real physical meaning of the Gaussian parameters, especially for the disc populations, is rather unclear.  \par
\begin{figure}

    \includegraphics[width=\columnwidth]{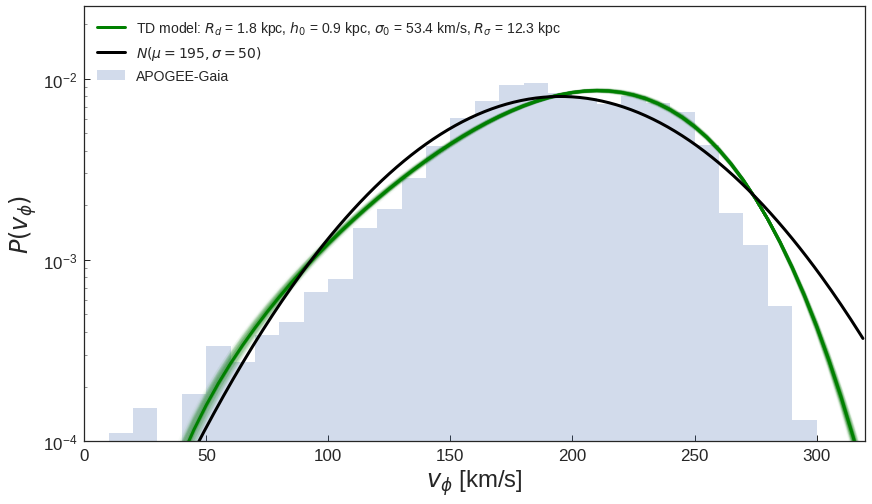}
    \caption{The distribution of $v_\phi$ for the thick disc. The solid histogram shows an APOGEE-Gaia sample (see text for details), while the green line shows our adopted model based on the work of \citet{schon2012}. In grey we show the canonical Gaussian distribution adopted by the Galaxia model ($\mu = 195$ km/s and $\sigma = 51$ km/s).}
    \label{vphis}
\end{figure}

In order to better depict the asymmetry in the azimuthal velocity distribution, \citet{schon2012} presented a new formula to describe the disc kinematics using physically meaningful parameters. It is constructed within a dynamical framework that connects the skewness of the $v_{\phi}$ distribution to the standard deviations of $v_R$ and $v_z$. Their model shows how the distribution of $v_{\phi}$ changes for slices in distance from the plane ($z$) and galactocentric radius ($R$). The final distribution, given in equation 29 of \citet{schon2012}, is a function of: the galactocentric radius of the measurement ($R_0$), the scale length of the observed population ($R_d$), the local radial velocity dispersion ($\sigma_0$), the scale length on which the radial velocity dispersion varies ($R_{\sigma}$) and the scale height of the vertical density profile ($h_0$). All the parameters reflects the characteristics of the dominant population, e.g. thin or thick disc stars, in a given sample. %
Our first update to the canonical Galaxia model is on the description of the thick disc kinematics. Instead of the canonical Gaussian distribution for the velocity components, we use the aforementioned physically motivated model described in \citet{schon2012} to draw the three velocity components of the mock data. To do this requires us to adopt values for the aforementioned parameters for the thick disc (i.e. $R_d$, $\sigma_0$, $R_{\sigma}$ and $h_0$). Since there is no firm consensus on the values of these parameters (see \citealt{bland} for a discussion), we obtain values using data from the APOGEE survey.
We take our data from the publicly available APOGEE-Gaia DR12  catalogue\footnote{\url{https://www.sdss.org/dr12/irspec/spectro_data/}}, selecting all stars within 1 kpc of the Sun with relative errors in parallax lower than 20\%, -1.1<[Fe/H]<-0.1 and 0.2<[Mg/Fe]<0.45. We note that our chemical selection is within the same range of the selected thick disc (alpha rich) stars from \citet{mackereth}.
We fit the APOGEE $v_\phi$ distribution according to \citet{schon2012}, using the Affine Invariant Markov Chain Monte Carlo \textit{emcee} package\footnote{\url{http://dfm.io/emcee/current/}} \citep{emcee}. In order to reduce degeneracies in the modelling we fix $R_d = 1.8$ kpc and $h_0 = 0.9$ kpc, leaving $\sigma_0$ and $R_{\sigma}$ free. This procedure gives $\sigma_0=53.4$ km/s and $R_{\sigma}=12.3$ kpc.
Although this is a simplistic approach, we are only concerned with obtaining a qualitatively good fit and have found that our results are not especially sensitive to the choice of parameters.
Fig. \ref{vphis} shows the resulting $v_{\phi}$ distribution from our model, compared to the observed APOGEE-Gaia thick disc sample. Our model provides a reasonable fit and one that is statistically superior to the canonical Gaussian model. Since the goal of this paper is not to find the best fit for the local thick disc, we defer more detailed modelling to future work.

\subsection{Stellar halo}\label{sec:modelhalo}
\citet{belo2018} have shown that, within 10 kpc, the stellar halo's velocity ellipsoid has a strong dependence on metallicity. As the metallicity increases, the anisotropy ($\beta$) of the halo also increases, indicating a radially biased population. Moreover, not only does the stellar halo have a mild rotation, the rotation signature is stronger for more metal poor stars. None of these features are incorporated into the original Galaxia stellar halo model, thus we updated the stellar halo model incorporating the rotation and anisotropy dependency on [Fe/H].
Since \citet{belo2018} only explore the halo for $z > 1$ kpc, we have extrapolated their results and assumed that the features observed for $1 < z < 3$ kpc are the same for $z <1 $ kpc. Table \ref{tab_param} shows the features of the updated stellar halo model, averaged over all [Fe/H]. Note that the tilt of the halo ellipsoid is likely to be negligible for our sample (see, for example, \citealt{wegg}) and so we are safe to neglect any correlations in the velocities.

\subsection{Fitting procedure}\label{sec:fitting}
Once the functions that describe the velocity distributions for our halo and thick disc populations are chosen, our modelling is very straightforward. We first select all the stars in the mock catalogue that follow the magnitude and distance criteria from the original ADQL query. For the thick disc and halo stars we re-sample their velocities using our updated velocity parameters and recalculate their $v_t$. We then compare the $v_t$ distribution to the observed distribution over the range $v_t > 200$ km/s. To do this, we take bins of width 10 km/s and calculate the maximum likelihood using a standard $\chi^2$ technique, where we have two free parameters corresponding to the re-weighting factors for the thick disc ($c_{TD}$) and halo ($c_{h}$). We do not re-weight the thin disc because the fraction of thin disc stars in this $v_t$ range is negligible (< 0.3\%), as expected.\par
Our adapted Galaxia model is described by the following equation,
\begin{equation}
	\text{N}^{i}_{\rm model} = \mathcal{N} \left[c_h \cdot \text{N}^i_h + c_{TD}\cdot \text{N}^i_{TD} + \text{N}^i_{td}\right]\,,
\end{equation}
where, $N^i_{\rm model}$ is the total number of stars in the i-th bin, $N^i_h$/$N^i_{TD}$/$N^i_{td}$ are the number of halo/thick-disc/thin-disc stars in the original mock catalogue, $c_h$ and $c_{TD}$ are the aforementioned re-weighting parameters, and $\mathcal{N}$ is the normalisation. Note that in order to maintain a good fit over the whole velocity range, we normalise over the entire range of $v_t$, not just $v_t > 200$ km/s, using the following normalising factor,
\begin{equation}
\mathcal{N} = \left( \text{N}^{tot}_{td} +c_h \cdot \text{N}^{tot}_h + c_{TD}\cdot \text{N}^{tot}_{TD} \right) ^{-1}\,,
\end{equation}
where the superscript $tot$ is the total number of mock stars for each component in the original Galaxia model (i.e. within the distance and magnitude range of the observed data and for all $v_t$).\par
We then compare $N^i_{\rm model}$ to the corresponding bin from the observed Gaia distribution, $\text{N}^i_{\rm Gaia}$, using a standard $\chi^2$ maximum likelihood technique, assuming Poisson uncertainties,
\begin{equation}\label{eq_likeli2}
\mathcal{L} \propto \prod_i \exp{\left(\frac{\text{N}^{i}_{\rm Gaia} - \text{N}^i_{\rm model}}{\sqrt\text{N}^{i}_{er,\rm Gaia}}\right)^2 } \,.
\end{equation}
$\text{N}^i_{\rm Gaia}$ is also normalized by the total number of Gaia stars in the same selection criteria. $\text{N}^{i}_{er,\rm Gaia}$ is the Poison error in the i-th bin.\par
To identify the values of $c_h$ and $c_{TD}$ that maximise our Equation \ref{eq_likeli2} we again use \textit{emcee}, finding $c_h = 3.78 \pm 0.02 $ and $c_{TD} = 0.52 \pm 0.01$. If we return to Fig. \ref{vts} we can now compare our model (green line) to the observed data, where it is clear that our re-weighting has produced a significant improvement in the tail of the distribution compared to the standard Galaxia model. Moreover, even though our fit used only the available data for the tail, we can see that the model is in good agreement with the $v_t$ distribution as a whole.   \par
\subsection{Results}\label{sec:model_result}
We can use these correction factors to calculate the local halo and thick disc fractions within 1 kpc and for $G<17$, obtaining $f_h = 0.47 \%$ and $f_{TD} = 6.52 \%$. In comparison, the Galaxia values are $0.12\%$ for the stellar halo and $10.95\%$ for the thick disc.

The fact that our values differ from the Galaxia ones is not necessarily surprising, as can be seen from our discussion in Section \ref{sec:intro}. For the halo, recent studies seem to be converging on fractions of around $0.4$ to $0.5 \%$ (e.g. \citealt{juric}, \citealt{kordopatis}, \citealt{posti}). For the thick disc fraction, there is no firm consensus.
This is mainly due to the degeneracy with its scale height and the reliance on statistical methods to disentangle thin and thick disc samples.
Low values of $f_{TD}$ have been reported previously, e.g. \citet{just} who claim that a massive thick disc is inconsistent with the kinematics of nearby K and M dwarfs.
Finally, with these updated fractions, our model predicts that $\approx$ 13\% of the high $v_t$ Gaia sample has kinematics consistent with the thick disc. We note that this result is not sensitive to our assumed form for the thick disc velocity distribution. For example, if we take $R_{\sigma}$, $\sigma_{0}$ and $R_d$ from the dynamical studies listed in \citet{bland} table 5, or if we simply use the canonical Gaussian distribution from Galaxia, then the fraction stays within the range 12 -- 14 \%. Nevertheless, we still favour our assumed thick-disc model due to the improved fit to the APOGEE data (Fig. \ref{vphis}).

\section{Chemistry and HR diagram}\label{sec:chem}
In the previous section we have shown that a kinematic decomposition of the Gaia high $v_t$ stars indicates that $\sim 13$ \% have thick disc kinematics, while the remaining $\sim 87$ \% have halo kinematics. This suggests that a dual population scenario is still plausible and so, to investigate this further, we will now explore the chemistry of these stars.

\subsection{LAMOST-Gaia high $v_t$}\label{sec:chem_lam}
The LAMOST spectroscopic survey telescope has been collecting low-resolution (R $\sim 2000$) data for millions stars since operations began in 2011 \citep{Luo2015}. This survey is ideal for our study, allowing us to analyse the chemistry of these high $v_t$ stars.
We take stellar parameters from the pipeline of \citet{boeche}, which has been updated to incorporate data from the fifth public data release\footnote{\url{dr5.lamost.org}}. After removing repeats, stars with signal-to-noise less than 40, and other artefacts, we cross-match with Gaia DR2 and obtain a final sample of 2982 stars with $v_t$ > 200 km/s.
For these stars we have full 6D phase-space, together with [Fe/H] and \alphaH. The median uncertainties on the chemistry are $\delta{\rm[Fe/H]}\approx 0.06$ dex and $\delta{[\alpha/Fe]}\approx 0.13$ dex. Although the uncertainties on \alphaH\, are too large to provide insights for individual stars, \citet{boeche} show that the systematics are small and hence {\alphaH} can be used on a statistical basis.

\begin{figure}

    \includegraphics[width=\columnwidth]{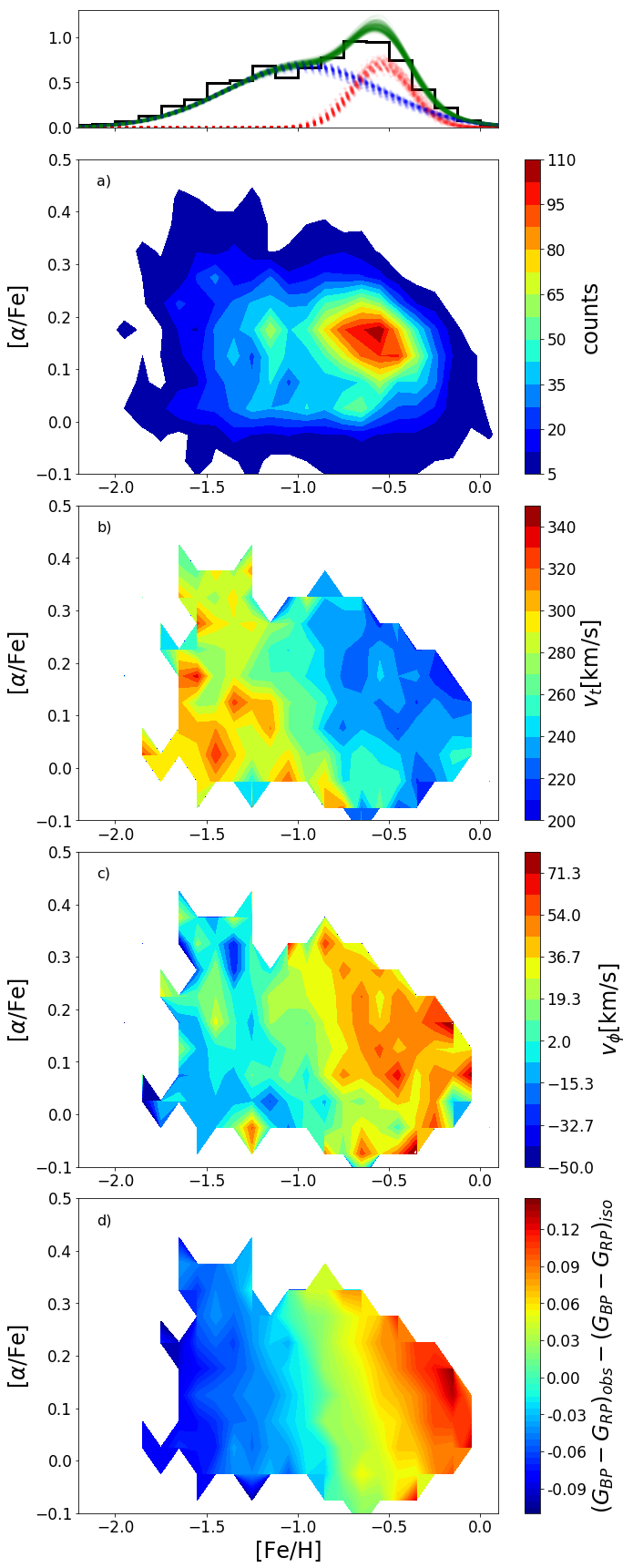}
    \caption{[Fe/H]-\alphaH\, plane for the Gaia stars with $v_t > 200$ km/s. The contour curves in each panel are colour-coded by a specific quantity: \textit{a)} number counts, \textit{b)} median $v_t$, \textit{c)} median $v_{\phi}$ and \textit{d)} median distance from the bisecting isochrone. We only show regions with at least 5 stars in each pixel, with each pixel being 0.1x0.05 dex. The top histogram shows the marginalised distributions of [Fe/H] and its double-Gaussian fit.}%
    \label{feh_lam}
\end{figure}

In Fig. \ref{feh_lam} we show the distribution of stars in the \alphaH-[Fe/H] plane, colour-coded by various properties. The density in this plane (Fig. \ref{feh_lam}\textit{a}) shows that there is a wide spread in metallicities and a hint of bimodality, with peaks at -1.2 and -0.6 dex. Neither the spread, nor the bimodality, would be expected for a pure stellar halo population. For example, although dealing with slightly larger volumes (within a few kpc), the studies of \citet{smith2009} and \citet{ivezic2008} found that the halo [Fe/H] distribution is well-fit by a single Gaussian with $\mu\approx-1.5$ and $\sigma\approx0.3$ and, in both cases, does not extent to solar metallicity. Furthermore, the more metal-rich stars have a relatively narrow peak in their \alphaH distribution, unlike the more metal-poor stars. This peak is what we would expect for a thick disc population \citep{hayden}.

Although a Gaussian distribution is unlikely to accurately represent the metallicity distribution for a complex system such as the halo or thick disc, for simplicity's sake we proceed to fit the distribution with two Gaussians. The blue and red lines in the top panel show the results of this fit, with parameters $\left(\mu_1 = -0.98; \sigma_1 = 0.41 ; w_1 = 0.71\right)$ and $\left(\mu_2 = -0.54; \sigma_2 = 0.17 ; w_2 =0.29\right)$, where $w_{i}$ is the normalisation factor for each Gaussian.

The more metal poor Gaussian is indicative of a halo population. It's broader than what is expected, extending to metallicities much higher than typical halo stars (e.g. \citealt{smith2009,ivezi}). The peak is also shifted slightly to higher metallicities. To check this further we have also analysed metallicities from the default LAMOST pipeline \citep[LASP,][]{Luo2015} and found that in this case the metal-poor stars peak around -1.22 dex, which is in better agreement with literature results for the nearby stellar halo. We note that the overall shape of the distribution, and the resulting double-Gaussian fit, is essentially unchanged, apart from this small shift and a slightly more prominent bimodality. \par
The second Gaussian is consistent with the expected [Fe/H] distribution for the canonical thick disc at low latitudes \citep{cheng}. Interestingly enough, it contributes about 30\% of the total distribution, which is significantly larger than the fraction predicted from the kinematic analysis ($\sim13\%$).
One should not over-interpret this simple decomposition; issues such as the overly broad halo distribution or the reduced thick-disc fraction are likely due to the fact that a double-Gaussian model does not accurately represent the underlying distributions. Whether the stellar halo is more metal rich for lower altitudes is still unclear and investigating this is beyond the scope of this work. A more realistic decomposition would consist of a narrower halo distribution, with a peak around -1.2 dex and spread extending to around -0.8 dex. The remaining stars would then be attributed to population with thick-disc chemistry. We return to this issue in more detail in Section \ref{sec:summary}.

Panels \textit{b} and \textit{c} in Fig. \ref{feh_lam} show the \alphaH-[Fe/H] plane colour-coded by the medians of $v_t$ and $v_{\phi}$, respectively. The differences between the metal rich and metal poor stars is clear, with a division around -0.7 dex: the former have significant rotation (with $v_{\phi} \gtrsim 35$ km/s), and much smaller $v_t$ (with $v_t \approx 200$ km/s); the latter have little rotation and larger values of $v_t$, typically $v_t>260$ km/s.
Thus, if one is interested in obtaining a clean sample of metal-poor stars with halo kinematics, a higher $v_t$ cut is recommended \citep[as advocated by, for example,][]{gould}. 

The value of net rotation for the metal rich stars is significantly slower than what we expect for the thick disc. This is because the cut of $v_t>200$ km/s removes most of the co-rotating stars, leaving a kinematically biased subset of thick disc stars. The cut also biases the halo population, removing any net rotation and shifting the mean towards lower values. Therefore the stars in our sample with weakly pro-grade motion are most-likely unrelated to the halo and are simply the low angular momentum tail of the thick disc population.

Despite the strong gradients with [Fe/H] in panels \textit{b} and \textit{c}, the trends with \alphaH\, are less clear. A weak trend can be found if one looks at fixed [Fe/H]. For example, for [Fe/H] = -0.8 dex it can be seen that median $v_{\phi}$ increases as \alphaH\, increases, as one would expect if the high \alphaH\, stars are dominated by thick-disc population. Similarly, at this [Fe/H] $v_t$ decreases as \alphaH\, increases, again reflecting the expected trend. These trends support the hypothesis that the accreted halo stars, as in \citep{hawkins}, are expected to be $\alpha$-poor compared to stars born in the Milky Way.

\subsection{The Blue and Red Sequences}\label{sec:bsrs}
As mentioned in Section \ref{sec:intro}, the intriguing feature of the high $v_t$ Gaia sample is the two distinct tracks in the HRD, which suggests the presence of two distinct populations. In order to better understand this, we now analyse the properties of the stars subdivided into two groups corresponding to the blue and red sequences (hereafter, BS and RS, respectively).

\begin{figure}
 	\includegraphics[width=\columnwidth]{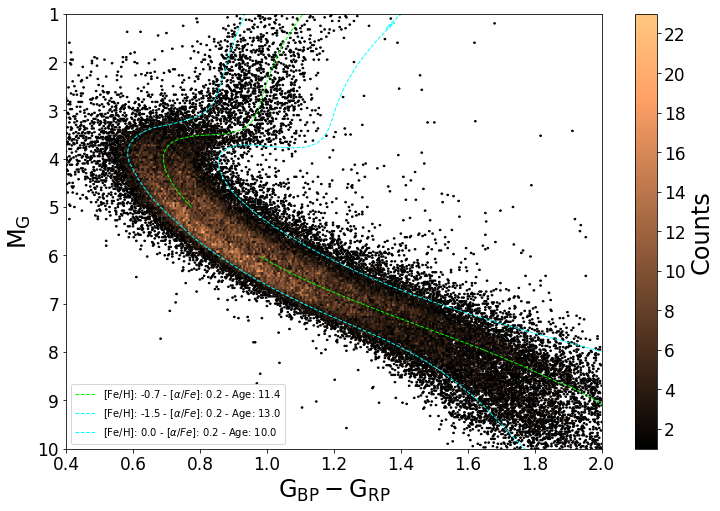}
    \caption{Dereddened Gaia HRD diagram for the high $v_t$ stars within 1 kpc.
    The green dashed line is the bisecting isochrone used to delineate the blue and red sequences. The cyan line shows two ``boundary" isochrones, illustrating the typical age and metallicity spread of the sample. }
    \label{mock_HRD}
\end{figure}

Fig. \ref{mock_HRD} shows the HRD after correcting for extinction using the \textit{MW.Combined15}\footnote{\url{https://github.com/jobovy/mwdust}} map from \citet{bovy_dust}.
From \citet{gallart} and \citet{sahlholdt}, we know that the two sequences
can be represented by isochrones with metallicities $\sim$-1.5 dex and and $\sim$-0.7 dex for BS and RS, respectively. Although we do not pursue a systematic fit of the HRD as in \citet{gallart}, we stress that between [Fe/H], \alphaH, mass and age, the former is the driving factor in producing the two clearly-separated sequences. One might wonder whether unresolved binaries could play a role in this double sequence, as equal mass binaries would cause a shift in magnitude by a factor $2.5\log2$, i.e. similar to the observed separation of the two sequences. However, the standard deviation of the epoch radial velocities\footnote{\url{http://gea.esac.esa.int/archive/documentation/GDR2/Gaia_archive/chap_datamodel/sec_dm_main_tables/ssec_dm_gaia_source.html}} for both sequences are very similar, implying that unresolved binaries are unlikely to play an important role in creating the double sequence.
In Fig. \ref{mock_HRD} we show that the two sequences are bifurcated about an isochrone of [Fe/H]=-0.7, \alphaH=0.2 and age=11.4 Gyrs. 

Fig. \ref{feh_lam}\textit{d} shows the \alphaH-[Fe/H] plane colour-coded by the colour-offset from the bisecting isochrone, i.e. ${\rm (BP-RP)_{obs} - (BP-RP)_{iso}}$. As expected, there is a direct correlation between colour and [Fe/H]. This means that a bimodality in [Fe/H], as we have seen in panel \textit{a}, will naturally result in a bimodality in colour. For the rest of this section we define RS/BS stars as those with colours redder/bluer than the bisecting isochrone. We obtain 31752 and 45355 stars in the RS and BS, respectively.

\begin{figure*}

    \includegraphics[scale=0.38]{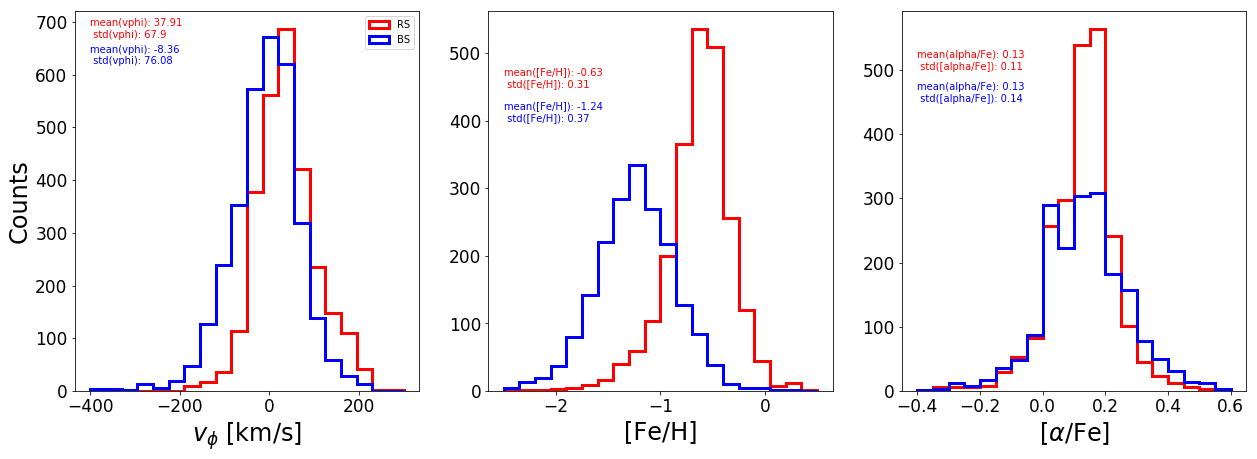}
    \caption{\textbf{Left panel:} \vphi\, distribution for the blue and red sequences in the HR diagram. The BS is consistent with a non-rotating population whereas the RS shows a significant sign of rotation. \textbf{Middle panel:} [Fe/H] histograms for both sequences. As expected, the RS is mainly composed of metal-rich stars, in contrast to the more metal-poor BS. \textbf{Right panel:} \alphaH\, histograms. The strong peak observed in the RS indicates the presence of stars with thick-disc chemistry.}
    \label{panel_HRD}
\end{figure*}

The left panel of Fig. \ref{panel_HRD} shows the $v_{\phi}$ histogram for both sequences. The BS is consistent with a population with zero rotation, whereas RS has a mild sign of rotation. The rotation signature in the RS is likely reflecting the fact that most of the thick disc contamination belongs to the RS. This contamination introduces some net rotation to this group of stars. As pointed out in the Section \ref{sec:chem_lam}, the fact that this sample is based on kinematic cut (i.e. $v_t > 200$ km/s) means that the inferred kinematic properties are going to be biased. For example, even though the BS stars have zero net rotation, we cannot infer that the stellar halo has zero net rotation because the pro-grade stars in this population are preferentially removed by the $v_t$ cut.

In order to quantify the contribution of stars with thick disc kinematics in the RS, we have applied our kinematic modelling to it, as described in Section \ref{sec:fitting}. We note that, as we normalise with the total number of stars in Gaia and in the mock catalogue, we do not need to know the number of RS/BS stars in the mock catalogue. Our fitting has only two free parameters, namely the re-weighting factors $c_h$ and $c_{TD}$. We find that the best fit to the $v_t$ distribution predicts that $\approx36\%$ of the RS sample has thick disc kinematics. The BS is consistent with having no stars with thick disc kinematics, and so the total fraction is $\approx16\%$, in agreement with Section \ref{sec:model_result}. This fraction is enough to produce the rotation signal observed in the RS.

The middle panel of Fig. \ref{panel_HRD} shows the metallicity distribution for both sequences. Given the direct correlation between metallicity and colour (Fig. \ref{feh_lam}\textit{b}), it is not surprising to find that the RS has more metal rich stars compared to the BS, although there is some overlap which is likely due to age effects or scatter from observational errors.
The mean metallicity of the RS ($\left<\rm{[Fe/H]}\right> = -0.63$) is higher than what we expect for the ``canonical'' halo, e.g. $-1.55$ dex \citep{smith2009} or $-1.45$ dex \citep{ivezi}, supporting the conjecture that there is thick-disc contamination in this sequence. As expected, the blue sequence ($\left<\rm{[Fe/H]}\right> = -1.24$) is more-consistent with a pure halo population. 

One could argue that the overlap in metallicities is caused by a poor choice of bisecting isochrone. However, we have tested it using a range of different isochrones and found that the results are similar. This spread can also be understood by looking at the bottom panel of Fig. \ref{feh_lam}, where the distance from the bisecting isochrone is slightly slanted, i.e. for the same median value ${\rm (BP-RP)_{obs}}-{\rm (BP-RP)_{iso}}$ there is an [Fe/H] gradient of about 0.2 dex. Therefore, even if we adopt a more metal rich/poor cut by shifting the bisecting isochrone slightly to the right/left, the RS would still have a tail towards the metal poor end.

Finally the \alphaH\, histograms, shown in the right panel of Fig. \ref{panel_HRD}, also provide clues to the differences between the two sequences. The RS distribution has a prominent peak, in contrast to the much broader BS distribution.
Although the halo should span a large range of \alphaH\, \citep[e.g.][]{deboer}, the thick disc should have a much narrower distribution \citep[e.g.][]{ishigaki}.
This can also be seen in figure 6 of \citet{dimatteo2018}. In this work they show that the canonical thick disc is located in the interval [Mg/Fe] = [0.25,0.35] dex, whereas the stars consistent with the stellar halo have a broader interval in [Mg/Fe] = [0.1,0.35] dex.
Therefore the \alphaH\, distributions for each sequence again suggest that the RS is a mix of stars with thick-disc and halo chemistry. 

\section{Dynamics}\label{sec:dyn}
We now investigate the orbits of stars in the high $v_t$ sample.  We calculate orbits using the \textit{Galpy} package\footnote{\url{http://github.com/jobovy/galpy}} \citep{galpy} with the axisymmetric potential \textit{MWPotential2014}.
Even though observational uncertainties and limitations in the adopted potential can affect our ability to recover true dynamical properties \citep[e.g.][]{coronado}, deeper insights can still be obtained.
An important orbital parameter to study is the eccentricity of a star. Each population of the Milky Way tends to have a different distribution in eccentricity, i.e., going from close to circular for the thin disc, to moderately eccentric for the thick disc and eccentric for halo stars.

\begin{figure}
    \includegraphics[scale=0.43]{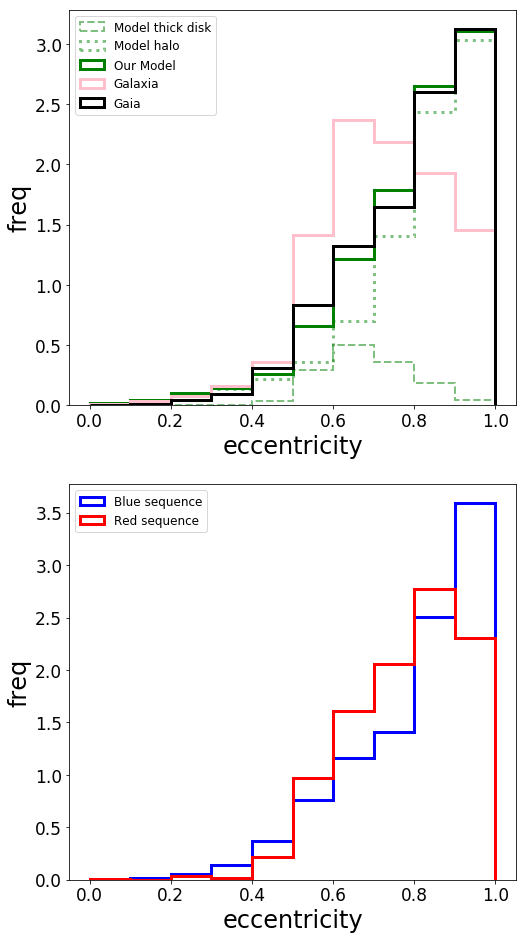}
    \caption{\textbf{Top panel: }The eccentricity distribution of the Gaia sample (black), Galaxia model (red) and our updated model (green). In contrast to the Galaxia canonical model, our new model provides a good match to the observed distribution. We also show the individual distributions for the mock thick disc (dashed) and halo (dotted). \textbf{Bottom panel:} Eccentricity distribution for the blue and red sequences. The RS appears to have more stars with moderate eccentricities, reflecting the presence of thick disc stars.}
    \label{dynpanel_ecc}
\end{figure}

Since the eccentricity distribution of the high $v_t$ sample will be biased due to the fact that it is based on a kinematic selection, we need to compare to model predictions. Fig. \ref{dynpanel_ecc} shows our observed eccentricity distribution (black), which is clearly not reproduced by the canonical Galaxia model (red).
On the other hand, our updated model (solid green) provides a very good agreement. The two main reasons for this improvement are: \textit{i.} the updated fractions of thick disc and halo stars in the sample and \textit{ii.} the new stellar halo model, which is more anisotropic compared to the canonical Galaxia stellar halo as advocated by \citet{belo2018}.
In the lower panel of Fig. \ref{dynpanel_ecc} we show the normalised histograms for the RS and BS eccentricity distribution. The RS has slightly fewer high eccentricity orbits compared to the BS. This  is due to the thick disc contamination, which (as can be seen in the top panel) peaks at an eccentricity of around 0.7. 

It has also been suggested that the orbital apocentre and maximum altitude can be used to infer the origin of these populations. In particular, \citet{haywood} noticed that the distribution in this space is not smooth and hence attributed it to the effects of an accretion event that heated the early disc. Indeed there is evidence that a major accretion event happened in the early galaxy \citep[e.g.][]{chiba, brook, helmi2018, belo2018}, which is likely to have some effect on the pre-existing disc \citep[e.g.][]{grand,jean}. However, as we show in Appendix \ref{sec:apo-zmax}, these structures in the apocentre and maximum altitude plane are likely just a consequence of the different orbital families that exist in the sample. The location of these structures are determined by the chosen potential under which the orbits are calculated, i.e. they should not be used to provide insights into the physical origins of the stars.

\section{Conclusion \& discussion}\label{sec:summary}
In this paper we have investigated the intriguing double-sequence in the HRD for high transverse velocity stars, as first shown in \citet{gaiahr}.
We have taken a different approach from previous studies in accounting for the expected thick-disk contamination, leading to a slightly different interpretation of the two sequences. In this section we bring together our results and discuss their implications.

\subsection{On kinematics}\label{sum_kin}
Our first step was to model the tail of the $v_t$, i.e., $v_t>200$ km/s. We have updated the description of the kinematics in the Galaxia stellar halo, using recent results from \citet{belo2018}. More importantly, we have also updated the thick disc kinematic model, adopting a physically meaningful azimuthal velocity distribution as described in \citet{schon2012}.
Our fitting procedure enables us to estimate the overall (i.e. for $v_t>0$) stellar halo and thick disc fractions within 1 kpc from the Sun, $f_h$ and $f_{TD}$ respectively.
We obtain a much better match to the tail (Fig. \ref{vts}) and find that $f_h = 0.47\%$ and $f_{TD}=6.52\%$.
The strength of this new approach is that it is based solely on kinematics, avoiding any pre-selection based on chemistry which could bias the result.

The best-fit model predicts that around $13\%$ of the high $v_t$ sample have thick-disc kinematics.
Therefore, if one is interested in selecting a pure halo sample based solely on a $v_t$ cut, we recommend a higher cut than 200 km/s. For example, a cut of $v_t > 220$ km/s produces a sample where $\approx 94\%$ stars have stellar halo kinematics, while a cut of 250 km/s has $\approx 98\%$. This is illustrated in Fig. \ref{feh_lam}, where we have shown that thick disc stars tend to have lower $v_t$ values. Adopting a low cut could produce biased samples and lead to erroneous conclusions regarding the nature of the stellar halo.

\subsection{On chemistry}\label{sec:summary_chem}
By cross-matching the high $v_t$ sample with LA\-MOST DR5 we have been able to investigate the chemistry of these stars (Section \ref{sec:chem_lam}). The [Fe/H] distribution appears bimodal, seemingly composed of a broad metal-poor component with halo-like chemistry and a narrower metal-rich component with thick-disk-like chemistry. 
Even though a Gaussian distribution may not provide a good representation of the [Fe/H] distribution for a given population, we fit our sample with a two-component Gaussian model and find that around $30\%$ of the stars belong to the latter (thick-disc-like) population.\par
As expected, the more metal-rich stars populate the RS and, as a consequence, the mean metallicity of the RS is higher than the ``canonical'' stellar halo.
There is some overlap in the [Fe/H] distributions of the two sequences, with the RS/BS having a tail towards the metal-poor/rich side, which is likely due to age effects or scatter from observational errors.\par
Our analysis has shown that there is a difference between the BS and RS \alphaH\, distributions, with the latter being consistent with a mix of thick disc and halo populations. This is in line with the findings of \citet{hayes}, who detected two distinct populations in a sample of metal-poor APOGEE stars, one with low ${\rm [Mg/Fe]}$ and one with high ${\rm [Mg/Fe]}$. They concluded that the low Mg stars are likely to be an accreted stellar halo population, due to the similarity of their chemistry with nearby dSph stars. On the other hand, the high Mg stars are chemically consistent with the metal poor end of the canonical thick disc, even though some of them have halo-like kinematics (see also \citealt{trincado}). 
This agrees with \citet{ishigaki}, who analysed a metal-poor sample of stars and found a difference in the abundance of $\alpha$ elements between the stellar halo and thick disc (also seen, e.g., in \citealt{nissen}, \citealt{hawkins}). Finally, \citet{bland2019} has also shown that the orbital angular momentum of $\alpha$-rich stars decreases with [Fe/H] (see their figure 4) indicating a transition between thick-disc and stellar halo.
\subsection{The metal rich counter-rotating stars}\label{sec:cr_feh}

\begin{figure}
 	\includegraphics[scale=0.43]{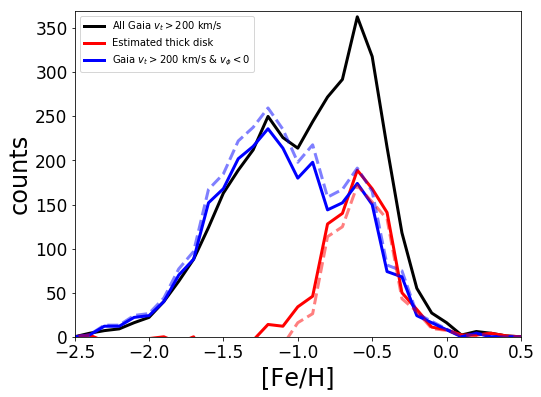}
    \caption{An investigation into the metallicity distribution of the high $v_t$ sample. The black line denotes the [Fe/H] distribution of the entire sample. The blue solid line is the metallicity distribution of all counter rotating stars ($v_{\phi}<0$) in the Gaia high $v_t$ sample, multiplied by 2. This provides an estimate of the halo distribution under the assumption that (a) the halo has no net rotation and (b) all thick-disc stars have prograde motion. The red solid curve is obtained by subtracting this halo-like (blue) distribution from the overall (black) distribution and should, in principle, mimic the thick disc metallicity distribution. The clear dual peak in the blue distribution indicates that there is a significant number of counter-rotating thick disc stars.
    The dashed lines uses a multiplication factor of $2.2$, which corresponds to the normalisation required to account for a rotation of $\sim 30$ km/s.}
    \label{cr_test}
\end{figure}

There is clearly a discrepancy in the above estimates for the amount of thick disc contamination in the high $v_t$ sample. The [Fe/H] distribution predicts around $30\%$ of the sample are consistent with the thick disc, while the kinematic decomposition suggests only $13\%$.
In order to further understand this, we can perform the following simple test:
if we assume, to first order, that the stellar halo has no net rotation and that the thick disc stars all have prograde motion, we can use the counter-rotating stars as a template for the metallicity distribution of the halo.
Given those assumptions, we can obtain the metallicity distribution of the thick disc by subtracting twice the counter-rotating population from the overall distribution. We present this in Fig. \ref{cr_test}. There are two interesting points to note: 
firstly, for the [Fe/H] distribution of the counter-rotating stars there appears to be a secondary peak coinciding with what we previously assumed to be the thick disc population (i.e. [Fe/H] = -0.6 dex); secondly, the subtracted histogram, which we expect to mimic the thick disc contribution, corresponds to $\approx 25\%$ of the data, which is still in disagreement with our kinematic modelling.
However, some studies have suggested that the stellar halo has prograde rotation and so we can account for this by multiplying the counter-rotating stars by a larger factor. For example, if we re-scale by 2.2, which would be the factor required to account for a prograde rotation of $\approx 30$ km/s, then the contribution of the thick disc is $\approx 18\%$. Even though this fraction is now in line with the kinematic modelling, it leaves us with the following curious question - of the stars that are kinematically classified as halo members, why do a non-negligible fraction have thick-disc-like chemistry?

This metal-rich peak in the retrograde distribution suggests that a number of thick-disc stars are, in fact, counter rotating. This is not intuitive, as one would expect that stars formed in a proto-galactic disc should have the same rotation direction. \citet{dimatteo2018} have already interpreted the presence of metal rich ([Fe/H]$\approx -0.5$) counter rotating stars as evidence for the heating of the early Milky Way disc by a merging satellite galaxy, such as the one proposed by \citet{belo2018,helmi2018}. However, it is also possible that some thick disc stars could be scattered onto counter rotating orbits during its formation. For example, \citet{clarke} shows that giant molecular clouds could have significantly heated the early disc, which may produce some counter-rotating stars (see \citealt{amarante} for further details). Another possibility is that some of these stars could be members of a different accretion event. For example the Sequoia system \citep{barba} has been shown to posses a number of retrograde metal rich globular clusters (such as NGC 3688; \citealt{myeong}) and hence these stars could be related to this event. \par
We can test the strength of this result and see how the distance estimate influences the number of retrograde and metal-rich ([Fe/H]>-0.7) stars in our sample. We adopt the parallax bias given by \citet{schon2019} and recalculate our distances and velocities. Although the introduced bias slightly shrinks the distances, we still find that the fraction of the retrograde metal-rich stars is roughly the same: 22.5\% (without parallax bias) and 21\% (including parallax bias).

\subsection{Nature of the Blue and Red Sequences}

We now bring together our findings in an attempt to clarify the nature of the BS and RS.
\begin{itemize}

    \item The BS is dominated by a halo-like component. Its properties match expectations for the Gaia-Sausage/Enceladus \citep{belo2018,helmi2018}, being radially anisotropic with little net rotation and having metallicities around $-2<$[Fe/H]$<-1$. 
    However, there is a tail of metal rich stars (e.g. $23\%$ have [Fe/H] > -1 dex), indicating that a small amount of chemical thick disc may be present. Our kinematic modelling suggests that all BS stars have halo-like kinematics, which implies these metal rich stars could be part of the aforementioned heated thick disc or the metal-rich tail of accreted material.
    
    \item The interpretation of the RS is slightly more complicated. Its chemistry suggests a division of $\sim20\%$ halo and $\sim80\%$ thick disc.
    On the other hand, the $v_t$ fit for this sequence suggests that around 64\% have ``halo-like" kinematics. These seemingly conflicting results can be resolved if we split this chemical thick disc material approximately evenly between a ``canonical" and a ``heated/accreted " thick disc, as discussed in the previous subsection. This is reinforced in Fig. \ref{cr_test} where we can see that the ``canonical'' and ``heated/accreted'' thick discs contribute approximately equally (assuming that the heated/accreted component has small net rotation). The $\sim 20\%$ stars with halo-like chemistry are radially anisotropic, matching expectations for the metal-rich part of the Gaia-Sausage/Enceladus \citep{belo2018,helmi2018}.

\end{itemize}

\subsection{Comparison to previous studies on the same sample}\label{sec:previous}

\citet{gallart} estimated the SFH for both sequences using a colour-magnitude diagram fitting technique, focusing on the turn-off region in a slightly larger volume than we have considered here ($\varpi > 0.5$ mas). They found that the RS and BS had similar SFHs (see fig. 2 of their paper)
and concluded that the former is made up of stars from the progenitor of the Milky Way and the latter from an accreted satellite galaxy. The pre-existing Milky Way would be more massive, and thus more metal rich, compared to the merging satellite. They estimated a 4:1 mass ratio between the merging satellite and our Galaxy's progenitor. Moreover, the RS (referred by them as in-situ halo) and the Milky Way's thick disc have a common origin. This picture suggests that a pre-existing disc was in place early in our Galaxy, certainly prior to the satellite's accretion.
\citet{dimatteo2018} also supports this scenario by showing that that $\approx 40\%$ of stars with [Fe/H] < -1 have thick disc like kinematics, similar to what we have found here. %
The study of \citet{sestito} provides further evidence, finding ultra-metal poor stars on thick-disc-like orbits. Since such metal-poor stars must be very old, if they were formed in the Milky Way then this supports the picture that the disc must have formed early (see e.g. \citealt{zolotov, brook2012}).
Our findings are similar to those of \citet{haywood}. They pointed out that the RS is likely to be made up mostly of thick disc stars and that the BS is likely coming from an accreted population, due to its \alphaH\, abundances.
We come to similar conclusions despite using a different technique, namely fitting of the kinematics, and have quantified the various fractions in each sequence.
Our studies are not in entire agreement, since we disagree with their interpretation of the patterns in the $R_{apo}-z_{max}$ plane. Rather than being a signature of the accretion event in the early galaxy, as we have pointed out in Section \ref{sec:dyn} and in Appendix \ref{sec:apo-zmax}, these features naturally occur due to the various orbital families that exist in the Galactic potential. Therefore, 
even though there is much evidence for such a merger event (e.g. \citealt{chiba}, \citealt{belo2018} and \citealt{helmi2018}), it is not required to explain the observed patterns in the $R_{apo}-z_{max}$ plane.
\section{Summary \& Final remarks} \label{sec:conclusion}

Our main findings from our analysis of the high tangential velocity stars ($v_t > 200$ km/s) in Gaia DR2 are listed below:
\begin{itemize}
    \item We estimate the fraction of stellar halo and (kinematic) thick disc within 1 kpc of the Sun to be $f_h = 0.47\%$ and $f_{TD} = 6.52\%$, respectively;
    
    \item This high $v_t$ sample has a contamination of 13\% stars with thick disc kinematics;

    \item Using LAMOST chemistry, we estimate that $\approx 30\%$ of high-$v_t$ stars have thick-disc like chemistry;
    
    \item There is a metal-rich population of counter-rotating stars, suggesting that a non-negligible number of thick-disc stars have retrograde orbits;
    
    \item All of the BS stars have stellar-halo like orbits. This suggests that the metal-rich tail of this sequence with [Fe/H] $> -1$ (i.e. $\sim$ a quarter of the stars) are likely to be part of the heated thick-disc or the metal-rich tail of accreted material;

    \item The RS, which contains mostly metal-rich stars, has a similar number of stars with halo and thick-disc kinematics, i.e. it is composed almost equally of both ``canonical" and ``heated" thick-disc components.

    \item Our interpretation of the two sequences is in alignment with \citet{haywood} and \citet{gallart}, namely the BS is mostly composed of accreted stars, whereas the RS is mostly composed of heated thick-disc material;

    \item The wedges in the $R_{apo}-z_{max}$ plane, which have been postulated to be a consequence of accretion \citep{haywood}, can be reproduced by a smooth (accretion-free) model such as Galaxia. These wedges are populated by different orbital families which exist in the adopted Galactic potential, i.e. one does not need to invoke accretion to explain their existence.

\end{itemize}

For a number of years the presence of multiple components in the stellar halo has been debated, but only now, thanks to more extensive observational data, is this picture taking form.
It is now known that the density of the stellar halo follows a broken power-law distribution, with a break at $\approx 20$ kpc (see e.g. \citealt{deason} and references therein). Within 20 kpc the stellar halo is dominated by a metal rich component related to the Gaia-Sausage/Enceladus event (\citealt{chiba}, \citealt{belo2018}, \citealt{helmi2018}), and the apocentres of these stars coincide with this break radius.
This merger has also left its imprints in the nearby stars, as we have shown throughout our paper. However, simulations have shown that kinematics alone are not enough to discriminate between accreted stars and those which were born in the main Milky Way progenitor and heated onto halo-like orbits, i.e. the ``in-situ" halo, (\citealt{zolotov}, \citealt{rodriguez-gomez}, \citealt{qu}).

We have shown that our updated fractions for the stellar halo and thick disc provide a good fit to the data, both in terms of the $v_t$ distribution (Fig. \ref{vts}) and the eccentricity distribution (Fig. \ref{dynpanel_ecc}).
However, if we look in detail we see that the chemical thick disc in the Gaia high $v_t$ sample include stars on retrograde orbits. Counter-rotating stars are not expected from standard kinematic models, such as the one we have adopted here \citep{schon2012}, and so alternative mechanisms such as additional heating or accretion are required. 
The properties of these stars are very uncertain and this clearly needs further study, ideally using detailed abundances and ages.
We note that while this paper was being refereed, \citet{belo2019} studied the chemistry and kinematics of stars likely heated during the Gaia-Sausage event (e.g. as seen also in \citet{haywood, dimatteo2018, mackereth}. They concluded that these stars, which they named ``Splash stars", are a distinct component of our Galaxy because their chemical and kinematic properties are different from either the accreted halo or canonical thick-disc. As a consequence this population could contain the first stars born in our Galaxy.
Simulations will also be extremely valuable to address the properties of the pre-existing disc and for understanding how this reacts to a significant merger event such as the Gaia-Sausage/Enceladus.

\section*{Acknowledgements}
The authors wish to thank the following people: Ralph Sch\"onrich for valuable assistance regarding his kinematic modelling procedure; Zhu Ling and Wyn Evans for advice regarding stellar orbits; John Vickers for general guidance; and the referee for useful comments that helped improve the clarity of this work.

J.A. and M.C.S acknowledge support from the National Key Basic Research and Development Program of China (No. 2018YFA0404501) and NSFC grant 11673083.
J.A. also acknowledges The World Academy of Sciences and the Chinese Academy of Sciences for the CAS-TWAS fellowship. 

This work has made use of data from the European Space Agency (ESA) mission {\it Gaia} (\url{https://www.cosmos.esa.int/gaia}), processed by the {\it Gaia} Data Processing and Analysis Consortium (DPAC,\url{https://www.cosmos.esa.int/web/gaia/dpac/consortium}). Funding for the DPAC has been provided by national institutions, in particular the institutions participating in the {\it Gaia} Multilateral Agreement.




\bibliographystyle{mnras}
\bibliography{main} 



\appendix \label{appendix}
\lstset{breaklines=true, language=SQL }
\section{ADQL query}\label{sec:adql}
The Gaia DR2 sample that we have used in this paper can be obtained with the following ADQL query.
\begin{lstlisting}
SELECT * FROM gaiadr2.gaia_source 
WHERE parallax_over_error > 10
AND parallax > 1
AND phot_g_mean_mag < 17
AND phot_g_mean_flux_over_error > 50
AND phot_rp_mean_flux_over_error > 20
AND phot_bp_mean_flux_over_error > 20
AND (sqrt(power(pmra,2) + power(pmdec,2)) * 4.74/parallax)>200
AND phot_bp_rp_excess_factor > 1. +0.015*power(bp_rp,2) 
AND phot_bp_rp_excess_factor < 1.3 +0.06**power(bp_rp,2)
AND visibility_periods_used > 8
AND astrometric_chi2_al/astrometric_n_good_obs_al - 5 < 1.44*max(1, exp(-0.4*phot_g_mean_mag-19.5))

\end{lstlisting}

\section{$apo-z_{max}$ plane}\label{sec:apo-zmax}
\begin{figure*}
    \includegraphics[scale=0.43]{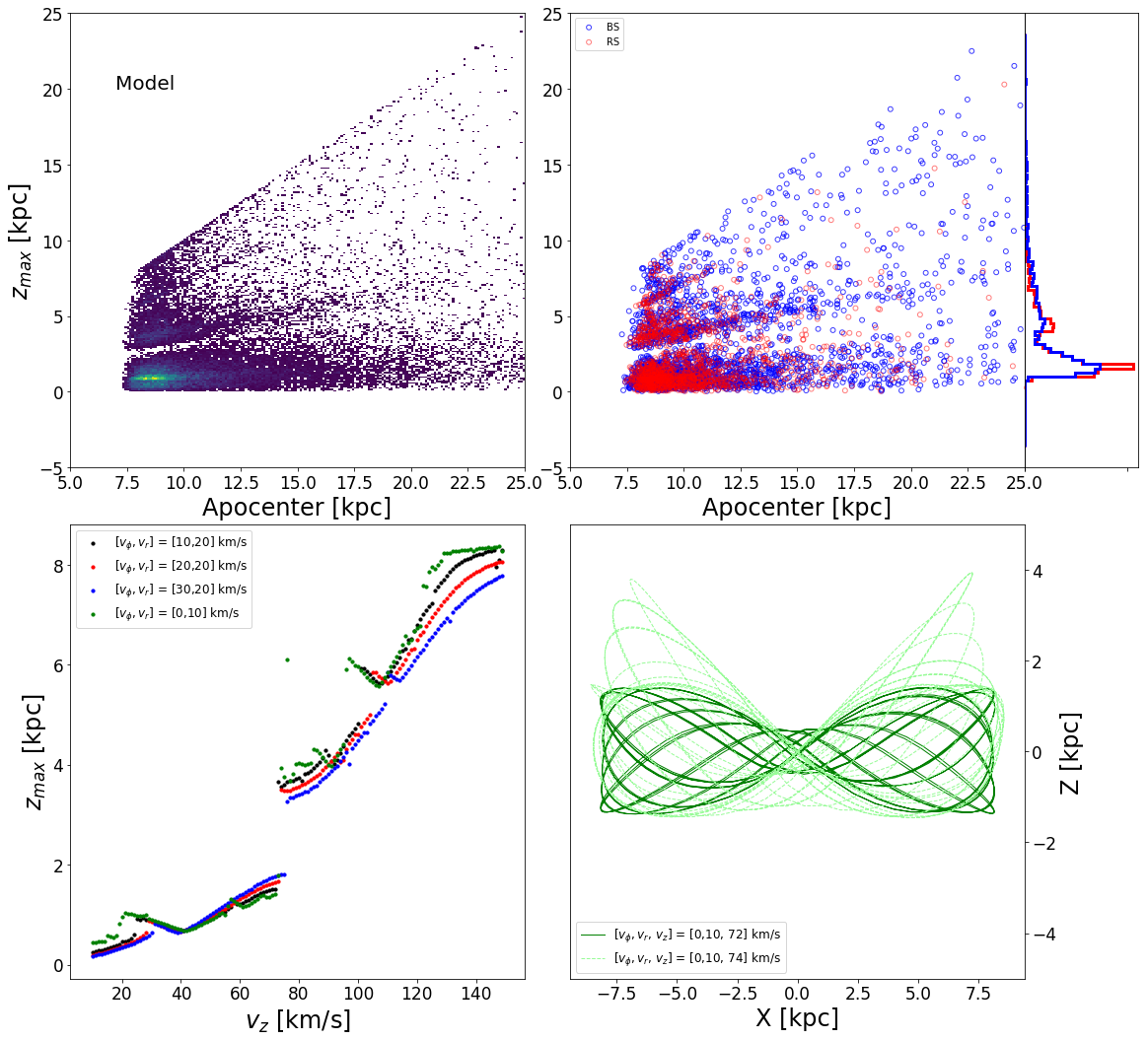}
    \caption{\textbf{Top row:} $R_{apo}-z_{max}$ plane for the mock stars from our updated model (left) and Gaia sample (right). The Gaia stars are colour-coded for the red and blue HRD sequences.
    The histogram to the right demonstrates that the BS and RS stars span a similar range of $z_{max}$.
    \textbf{Bottom row:} the left panel shows $z_{max}$ as a function of $v_z$ for \textit{MWPotential2014}, where $v_R$ and $v_{\phi}$ have been fixed to the values listed in the legend. The gaps are related to changes in orbital families. This is illustrated in the right panel, where we show two example orbits from either side of the gap at $v_z = 73$ km/s. Even though the change in $v_z$ for these two orbits is less than 3\%, there is a dramatic change in $z_max$ as the orbital family changes.}
    \label{dynpanel_apo}
\end{figure*}

As mentioned in Section \ref{sec:dyn}, \citet{haywood} investigated the distribution of high $v_t$ stars in the $R_{apo}-z_{max}$ plane. They noticed that the stars followed tracks in this plane and attributed this to the effects of accretion. We show the observed distribution in the top-right panel of Fig. \ref{dynpanel_apo}, where the stars have been split into the blue and red sequences. The ``lumpiness'' is clear, i.e. the stars seem to prefer discrete ``tracks" in this plane. Although the RS stars are concentrated towards lower $z_{max}$ and shorter apocenter, we note that neither sequence seems to follow a preferred track. This is confirmed in the top right histogram of Fig. \ref{dynpanel_apo}, where we can see that both sequences have extended distributions towards higher $z_{max}$ and have peaks at the same values of $z_{max}$.

It is known that halo stars can be clustered in a characteristic energy-Jacobi diagram due to resonant trapping, e.g. as in \citet{moreno}.
Inspired by this fact, we now perform a couple of simple tests to determine if the observed tracks can be explained by resonant effects with an axisymmetric Galactic potential.

Using our updated model with new values for $f_{TD}$, $f_{h}$ and new stellar kinematics (see Section \ref{sec:model} for details), we draw a sample of mock stars within 1 kpc and with $v_t>200$km/s. We calculate their orbits and plot the resulting distribution in the $R_{apo}-z_{max}$ plane in the top left panel of Fig.\ref{dynpanel_apo}. This mock distribution shows the same tracks as the observed data. The fact that we observe these tracks in our mock data, despite the smooth underlying kinematics with no accretion, tells us that this is simply a natural consequence of resonant effects.

We illustrate this further in the bottom-left panel of Fig. \ref{dynpanel_apo}. Here we show how $z_{max}$ varies as a function of $v_z$ for a star initially at the Sun's position. To simplify the situation, we fix the values for the radial and azimuthal velocities. As expected, $z_{max}$ increases with $v_z$, but this is not a 1:1 trend; there are both discontinuities and regions where the trend is reversed. For example, as $v_z$ approaches $\sim80$ km/s we can see $z_{max}$ ``jump'' from $\sim2$ kpc to $\sim3.5$ kpc. This corresponds to the significant gap seen at $\sim 2.5$ kpc in the upper panels, both for the observed and mock data.
The nature of this gap is illustrated in the bottom-right panel, where we have shown an example orbit from either side of this discontinuity. Both orbits have the same values for $v_R$ and $v_{\phi}$, but have $v_z = 72$ and 74 km/s. Even though the change in $v_z$ is less than 3\%, there is a dramatic change in $z_max$ as the orbital family changes. 
Therefore it is clear that the tracks observed in the $R_{apo}-z_{max}$ plane are due to transitions from one orbital family to another. The locations of these gaps are determined by the adopted potential and cannot be used to learn about the physical origins of the stars. 
%




\bsp	
\label{lastpage}
\end{document}